\documentclass[a4,twocolumn, superscriptaddress]{revtex4}
\usepackage{amsmath,amsthm,amsfonts,amssymb}
\usepackage{graphicx}

\usepackage{color}

\usepackage{mciteplus}

\renewcommand{\mcitesubrefform}{[\arabic{mcitebibitemcount}]\alph{mcitesubitemcount}} 
\begin{document}

\mciteSetSublistMode{f}




\textbf{Adams and Olmsted Reply:} Wang \cite{WangComment} makes the
following points about our Letter \cite{adams:067801}: (1) He infers
that, ``contrary to its title, shear banding [in \cite{adams:067801}]
emerged from monotonic curves only if there was a stress gradient'',
and he points out that nonquiescent relaxation was found
(experimentally) after step strain in geometries \emph{without} a
stress gradient \cite{BWW2009}. (2) He disagrees with the values of
the parameters we used. (3) In some recent experiments the flow was
homogeneous after cessation of step strain, and only
\textit{subsequently} developed nonquiescent macroscopic motion
\cite{BWW2009}. We only showed step strains that developed an
inhomogeneity \textit{before} cessation of flow, as in
\cite{wang:187801}.

\begin{figure}[htb]
\includegraphics[width = 0.38\textwidth]{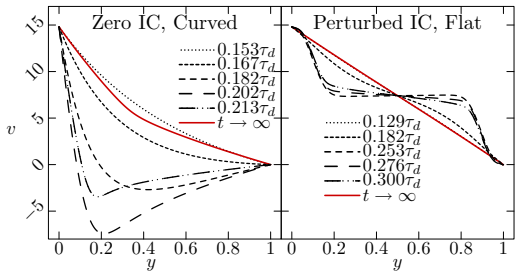}
\caption{(color online) Startup transients for (left) a cone angle of
  $4^{\circ}\;\;(q = 2\times 10^{-3})$ ; and (right) a flat geometry
  with noisy initial polymer shear stress $\Sigma_{xy}(0)$ of a few
  percent, with $\dot{\gamma}\tau_d = 14.8$, $\beta = 0.728$ and
  $\epsilon = 10^{-5}$, and $\tau_d/\tau_R = 10^3$. }
\label{fig:transients}
\end{figure}

(1) As our title stated \cite{adams:067801}, we showed that a fluid
with a monotonic constitutive curve based on Doi-Edwards (DE) theory
can have signatures similar to shear banding. These signatures arise
from a stress gradient (\textit{e.g.} the bowed steady state velocity
profile obtained in the stress gradient of a cone and plate rheometer
\cite{tapadia:016001} or transient banding-like profiles during
startup).  Flat geometries can have \textit{transient} banding-like
signatures: \textit{e.g.}  two clearly defined bands of shear rates
during large amplitude oscillatory shear (LAOS)
\cite{adams:067801,hu07,*tapadia:196001, *zhou:189}, or inhomogeneous
banding-like transients during startup flows in presence of
inhomogeneous spatial fluctuations (noise) (Fig.~\ref{fig:transients})
\cite{adams:067801}.

\vspace{-2ex}

\vspace{-2ex}


(2) Our parameters were matched to experiment, for a nonlinear model
in which the parameters $\tau_d$ and $\tau_R$ roughly correspond to
their rigorously defined counterparts in linear rheology. Because we
use (the best available) crude nonlinear theory, the parameters do not
correspond precisely.  We used $\epsilon = \eta/(G \tau_d) \simeq
10^{-5}$ based a plateau modulus $G\simeq3\textrm{kPa}$, reptation
time $\tau_d\simeq 20\,\textrm{s}$, and solvent viscosity
$\eta\simeq1\textrm{Pa s}$ \cite{ravindranath2008}. Although
$\tau_d/\tau_{R}\sim 10^3$ implies too many entanglements, it fits the
experimental nonlinear rheology well \cite{tapadia:016001}. This
inconsistency is an unsatisfactory feature of current theory.

\begin{figure}[!h]
\includegraphics[width = 0.38\textwidth]{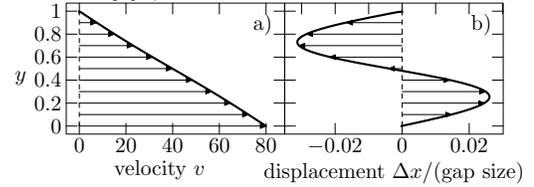}
\caption{Recoil displacement (b) at $t=
  0.08\tau_d$ after cessation of homogeneous shear (a) at
  $t_\textrm{stop} = 0.0375 \tau_d$ with $\dot{\gamma} \tau_d = 80$,
  ($\gamma = 3$) for the Rolie-Poly model with $\beta = 0.728$,
  $\epsilon = 10^{-5}$ and $q = 0$, with initial noise.}
\label{fig:params}
\vspace{-5ex}
\end{figure}

(3) The step strain results in \cite{adams:067801} should be compared
with \cite{wang:187801}-(Fig.~5), where the velocity profile became
inhomogeneous before cessation. 
Fig.~\ref{fig:params} shows a calculation in which inhomogeneities
develop only \textit{after} cessation of flow, during a strong recoil.
This is for startup in a flat geometry, with noisy initial conditions,
and resembles \cite{wang:187801}-(Fig.~3) if there were no
experimental wall slip.

Wang's newest experiments show dramatic rupture and internal fracture,
despite a homogeneous velocity before cessation \cite{BWW2009}
(similar fracture planes could be interpreted in
\cite{WangMM2007}-(Fig. 3f), but in a cone-and-plate geometry;
moreover, those data are also consistent with wall slip and simple
recoil). Our calculations (Fig.~\ref{fig:params}) go some way towards
modelling this phenomenon, but do not capture this rupture, and have
not yet been adequately modified to incorporate slip. It remains a
strong challenge to distinguish which experimental features are
captured by tube models, and which (\textit{e.g.}  rupture) require
new physical insight. One suggestion is the ``elastic yielding'' in
\cite{WangComment} which may be similar to modifying the DE model to
incorporate the instability of the spatial distribution of
entanglements \cite{degennes07}. In fact, the instability in the DE
model occurs when the shear rate greatly exceeds the reptation time,
which is one criterion for elastic yielding postulated in
\cite{WangComment}.

\vspace{0.5ex}

J. M. Adams$^{1}$ and P. D. Olmsted$^{2}$

$^{1}$Department of Physics, University of Surrey, 

Guildford, GU2 7XH, United Kingdom

$^{2}$School of Physics \& Astronomy, University of Leeds

Leeds, LS2 9JT, United Kingdom\\

\vspace{-1.5ex}

Recieved 9 October 2009

DOI: 10.1103/PhysRevLett.103.219802


\vspace{-3.5ex}


\begin{mcitethebibliography}{11}

\expandafter\ifx\csname natexlab\endcsname\relax\def\natexlab#1{#1}\fi
\expandafter\ifx\csname bibnamefont\endcsname\relax
  \def\bibnamefont#1{#1}\fi
\expandafter\ifx\csname bibfnamefont\endcsname\relax
  \def\bibfnamefont#1{#1}\fi
\expandafter\ifx\csname citenamefont\endcsname\relax
  \def\citenamefont#1{#1}\fi
\expandafter\ifx\csname url\endcsname\relax
  \def\url#1{\texttt{#1}}\fi
\expandafter\ifx\csname urlprefix\endcsname\relax\def\urlprefix{URL }\fi
\providecommand{\bibinfo}[2]{#2}
\providecommand{\eprint}[2][]{\url{#2}}
\vspace{-2ex}
\bibitem[{\citenamefont{Wang}(2009)}]{WangComment}
\bibinfo{author}{\bibfnamefont{S.-Q.} \bibnamefont{Wang}},
  \bibinfo{journal}{Phys. Rev. Lett.} \textbf{\bibinfo{volume}{XX}},
  \bibinfo{pages}{XXX} (\bibinfo{year}{2009})\relax
\mciteBstWouldAddEndPuncttrue
\mciteSetBstMidEndSepPunct{\mcitedefaultmidpunct}
{\mcitedefaultendpunct}{\mcitedefaultseppunct}\relax
\EndOfBibitem
\bibitem[{\citenamefont{Adams and Olmsted}(2009)}]{adams:067801}
\bibinfo{author}{\bibfnamefont{J.~M.} \bibnamefont{Adams}} \bibnamefont{and}
  \bibinfo{author}{\bibfnamefont{P.~D.} \bibnamefont{Olmsted}},
  \bibinfo{journal}{Phys. Rev. Lett.} \textbf{\bibinfo{volume}{102}},
  \bibinfo{eid}{067801} (\bibinfo{year}{2009})\relax
\mciteBstWouldAddEndPuncttrue
\mciteSetBstMidEndSepPunct{\mcitedefaultmidpunct}
{\mcitedefaultendpunct}{\mcitedefaultseppunct}\relax
\EndOfBibitem
\bibitem[{\citenamefont{Boukany et~al.}(2009)\citenamefont{Boukany, Wang, and
  Wang}}]{BWW2009}
\bibinfo{author}{\bibfnamefont{P.~E.} \bibnamefont{Boukany}},
  \bibinfo{author}{\bibfnamefont{S.-Q.} \bibnamefont{Wang}}, \bibnamefont{and}
  \bibinfo{author}{\bibfnamefont{X.~R.} \bibnamefont{Wang}},
  \bibinfo{journal}{Macromolecules} \textbf{\bibinfo{volume}{42}},
  \bibinfo{pages}{6261} (\bibinfo{year}{2009})\relax
\mciteBstWouldAddEndPuncttrue
\mciteSetBstMidEndSepPunct{\mcitedefaultmidpunct}
{\mcitedefaultendpunct}{\mcitedefaultseppunct}\relax
\EndOfBibitem
\bibitem[{\citenamefont{Wang et~al.}(2006)\citenamefont{Wang, Ravindranath,
  Boukany, Olechnowicz, Quirk, Halasa, and Mays}}]{wang:187801}
\bibinfo{author}{\bibfnamefont{S.-Q.} \bibnamefont{Wang}},
  \bibnamefont{et~al.}, \bibinfo{journal}{Phys. Rev. Lett.}
  \textbf{\bibinfo{volume}{97}}, \bibinfo{eid}{187801}
  (\bibinfo{year}{2006})\relax
\mciteBstWouldAddEndPuncttrue
\mciteSetBstMidEndSepPunct{\mcitedefaultmidpunct}
{\mcitedefaultendpunct}{\mcitedefaultseppunct}\relax
\EndOfBibitem
\bibitem[{\citenamefont{Tapadia and Wang}(2006)}]{tapadia:016001}
\bibinfo{author}{\bibfnamefont{P.}~\bibnamefont{Tapadia}} \bibnamefont{and}
  \bibinfo{author}{\bibfnamefont{S.-Q.} \bibnamefont{Wang}},
  \bibinfo{journal}{Phys. Rev. Lett.} \textbf{\bibinfo{volume}{96}},
  \bibinfo{eid}{016001} (\bibinfo{year}{2006})\relax
\mciteBstWouldAddEndPuncttrue
\mciteSetBstMidEndSepPunct{\mcitedefaultmidpunct}
{\mcitedefaultendpunct}{\mcitedefaultseppunct}\relax
\EndOfBibitem
\bibitem[{\citenamefont{Hu et~al.}(2007)\citenamefont{Hu, Wilen, Philips, and
  Lips}}]{hu07}
\bibinfo{author}{\bibfnamefont{Y.~T.} \bibnamefont{Hu}}, \bibnamefont{et~al.},
  \bibinfo{journal}{J. Rheol} \textbf{\bibinfo{volume}{51}},
  \bibinfo{pages}{275} (\bibinfo{year}{2007})\relax
\mciteBstWouldAddEndPuncttrue
\mciteSetBstMidEndSepPunct{\mcitedefaultmidpunct}
{\mcitedefaultendpunct}{\mcitedefaultseppunct}\relax
\EndOfBibitem
\bibitem[{\citenamefont{Tapadia et~al.}(2006)\citenamefont{Tapadia,
  Ravindranath, and Wang}}]{tapadia:196001}
\bibinfo{author}{\bibfnamefont{P.}~\bibnamefont{Tapadia}},
  \bibinfo{author}{\bibfnamefont{S.}~\bibnamefont{Ravindranath}},
  \bibnamefont{and} \bibinfo{author}{\bibfnamefont{S.-Q.} \bibnamefont{Wang}},
  \bibinfo{journal}{Phys. Rev. Lett.} \textbf{\bibinfo{volume}{96}},
  \bibinfo{eid}{196001} (\bibinfo{year}{2006})\relax
\mciteBstWouldAddEndPuncttrue
\mciteSetBstMidEndSepPunct{\mcitedefaultmidpunct}
{\mcitedefaultendpunct}{\mcitedefaultseppunct}\relax
\EndOfBibitem
\bibitem[{\citenamefont{Zhou et~al.}(Monterey, CA, AIP,
  2008)\citenamefont{Zhou, Ewoldt, Cook, and McKinley}}]{zhou:189}
\bibinfo{author}{\bibfnamefont{L.}~\bibnamefont{Zhou}}, \bibnamefont{et~al.},
  \bibinfo{journal}{Proceedings of the XVth {I}nternational {C}ongress on
  {R}heology} p. \bibinfo{pages}{189} (\bibinfo{year}{Monterey, CA, AIP,
  2008})\relax
\mciteBstWouldAddEndPuncttrue
\mciteSetBstMidEndSepPunct{\mcitedefaultmidpunct}
{\mcitedefaultendpunct}{\mcitedefaultseppunct}\relax
\EndOfBibitem
\bibitem[{\citenamefont{Ravindranath and Wang}(2008)}]{ravindranath2008}
\bibinfo{author}{\bibfnamefont{S.}~\bibnamefont{Ravindranath}}
  \bibnamefont{and} \bibinfo{author}{\bibfnamefont{S.-Q.} \bibnamefont{Wang}},
  \bibinfo{journal}{J. Rheol.} \textbf{\bibinfo{volume}{52}},
  \bibinfo{pages}{957} (\bibinfo{year}{2008})\relax
\mciteBstWouldAddEndPuncttrue
\mciteSetBstMidEndSepPunct{\mcitedefaultmidpunct}
{\mcitedefaultendpunct}{\mcitedefaultseppunct}\relax
\EndOfBibitem
\bibitem[{\citenamefont{Ravindranath and Wang}(2007)}]{WangMM2007}
\bibinfo{author}{\bibfnamefont{S.}~\bibnamefont{Ravindranath}}
  \bibnamefont{and} \bibinfo{author}{\bibfnamefont{S.-Q.} \bibnamefont{Wang}},
  \bibinfo{journal}{Macromolecules} \textbf{\bibinfo{volume}{40}},
  \bibinfo{pages}{8031} (\bibinfo{year}{2007})\relax
\mciteBstWouldAddEndPuncttrue
\mciteSetBstMidEndSepPunct{\mcitedefaultmidpunct}
{\mcitedefaultendpunct}{\mcitedefaultseppunct}\relax
\EndOfBibitem
\bibitem[{\citenamefont{{de}~Gennes}(2007)}]{degennes07}
\bibinfo{author}{\bibfnamefont{P.~G.} \bibnamefont{{de}~Gennes}},
  \bibinfo{journal}{Eur. Phys. J. E} \textbf{\bibinfo{volume}{23}},
  \bibinfo{pages}{3} (\bibinfo{year}{2007})\relax
\mciteBstWouldAddEndPuncttrue
\mciteSetBstMidEndSepPunct{\mcitedefaultmidpunct}
{\mcitedefaultendpunct}{\mcitedefaultseppunct}\relax
\EndOfBibitem
\end{mcitethebibliography}
\ifx\mcitethebibliography\mciteundefinedmacro
\PackageError{apsrevM.bst}{mciteplus.sty has not been loaded}
{This bibstyle requires the use of the mciteplus package.}\fi

\end{document}